\documentclass[3p,times,twocolumn]{elsarticle}

\usepackage{amssymb}

\usepackage[figuresright]{rotating}

\newcommand{\beq}{\begin{equation}}
\newcommand{\eeq}{\end{equation}}
\newcommand{\beqa}{\begin{eqnarray}}
\newcommand{\eeqa}{\end{eqnarray}}

\def\0{{\bf 0}}

\begin{document}

\begin{frontmatter}


\title{Heavy flavor in nucleus-nucleus and proton-nucleus:\\ quenching, flow and correlations} 

\author[a]{Marzia Nardi\corref{mn}}
 \ead{nardi@to.infn.it}
 \cortext[mn]{Corresponding author}
\address[a]{INFN - Sezione di Torino, via P.Giuria 1, 10125 Torino, Italy}

\author[a]{A. Beraudo}
\author[a]{A. De Pace}
\author[a]{M. Monteno}
\author[a]{F. Prino}

\begin{abstract}
We present recent results for  heavy-flavor observables in  nucleus-nucleus collisions at LHC energies, obtained
with the POWLANG transport setup. 
The initial creation of  $c\bar{c}$ and $b\bar{b}$
pairs is simulated with a perturbative QCD approach (POWHEG+PYTHIA); their propagation in the medium (created in the nucleus-nucleus or in proton-nucleus collision) is  studied with the relativistic Langevin equation, here solved using weak-coupling transport coefficients.
 Successively, the heavy quarks hadronize in the medium. We compute the nuclear modification factor
 and the elliptic flow parameter of the final $D$ mesons both in nucleus-nucleus and in (for the first time, in the POWLANG setup) proton-nucleus collisions
and compare our results to experimental data. 
\end{abstract}


\end{frontmatter}



\section{Introduction}

The purpose of our work is to provide a comprehensive setup for the study of
heavy-flavor observables in high-energy hadronic and nuclear
collisions, from the  $Q\overline{Q}$ production in hard partonic processes to the detection in the experimental apparatus.

In a series of papers~\cite{lange0,lange1,lange2,lange3} over the last few years we developed a complete setup (referred to as POWLANG) for the study of heavy flavour observables in high-energy nucleus-nucleus (AA) collisions, describing the initial hard production of the $Q\overline{Q}$ pairs and the corresponding parton-shower stage through the POWHEG-BOX package~\cite{POW,POWBOX} and addressing the successive evolution in the plasma through the relativistic Langevin equation.
 Here, following Ref. \cite{lange3}, we supplement our numerical tool by modeling the hadronization of the heavy quarks accounting for the presence of a surrounding medium made of light thermal partons feeling the collective flow of the local fluid cell. Moreover, we present our first (preliminary) results for proton-nucleus (pA) collisions.

\section{Heavy flavour in proton-proton collisions}

Because of their large mass, the initial production of $c$ and $b$ quarks is a
short-distance process described by perturbative QCD (pQCD). 
 For this purpose we rely on a standard pQCD public tool, namely POWHEG-BOX, in which the hard $Q\overline{Q}$ 
event is interfaced with a shower
stage described by PYTHIA~\cite{PYTHIA}, to include the effects of Initial- and
Final-State Radiation~\cite{lange0,lange1}.

Experimental data obtained in $pp$ collisions can be exploited to validate the theoretical calculations used to simulate the initial hard $Q\overline{Q}$ production.  
In our setup, the heavy quarks are created in pairs by the POWHEG-BOX event generator. 
Eventually, heavy-quark hadronization and the final decays of the $D$ ($B$) mesons are simulated with PYTHIA, which is also used to describe the parton shower stage.

\begin{figure}
\begin{center}
\includegraphics[clip,width=0.44\textwidth]{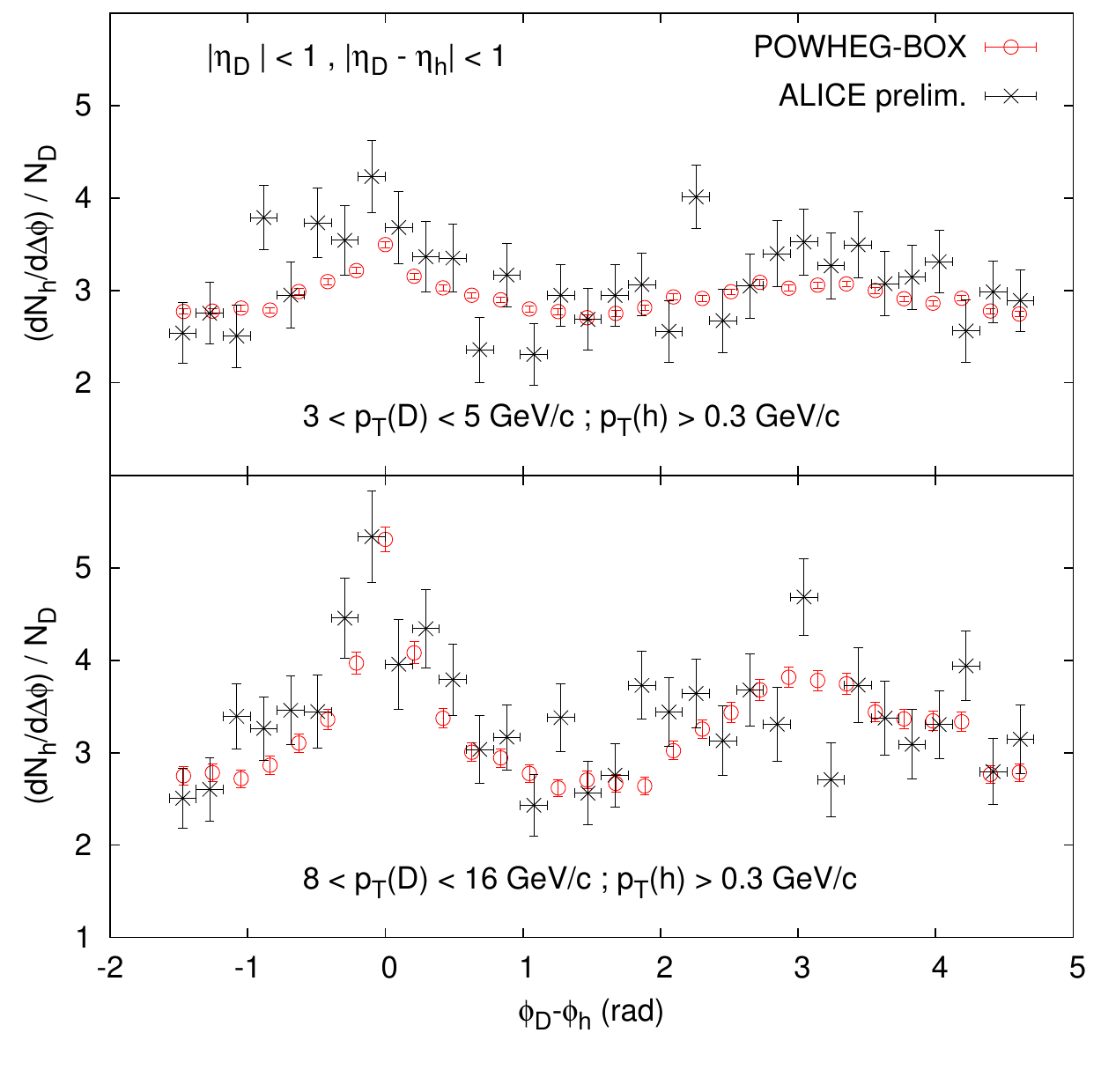}  
\caption{Azimuthal $D\!-\!h$ correlations in pp collisions at $\sqrt{s}\!=\!7$ TeV for different  $p_T$-cuts compared to preliminary ALICE data~\cite{sandro,rossi}.}\label{fig:ppcorr} 
\end{center}
\end{figure}

In Fig.~\ref{fig:ppcorr} we show our results for $D\!-\!h$ azimuthal correlations compared to preliminary ALICE data~\cite{sandro,rossi}, for two different $p_T$-intervals of the charmed meson. In our simulation $D^0$, $\bar{D}^0$ and $D^\pm$ are used as trigger particles and the light hadrons are limited to charged pions and kaons, protons and antiprotons, excluding the weak decays of $\Lambda$ and $K^0$. Any $D$-meson is correlated with all the light hadrons (except its own decay products) created in the same event.
The near-side peak takes contribution both from correlations present at the partonic level (from $Q\overline{Q}$ pairs arising from gluon splitting) and from hadrons coming from the fragmentation of the same string of the parent heavy quark. Our results include also the simulation of the Underlying Event due to Multiple Parton Interactions, performed with PYTHIA 6.4, which gives rise to the pedestal observed in Fig.~\ref{fig:ppcorr}.

\section{Heavy flavour in pA and AA collisions: setups}

In nuclear collisions  the $c\bar{c}$ or $b\bar{b}$ production given by the pQCD calculation must be modified in two ways: $i)$ the nuclear parton distribution functions should be corrected for shadowing or antishadowing effects 
(we have adopted here the
EPS09 scheme~\cite{eps}); $ii)$ the colliding partons acquire, on average, a larger
transverse momentum during the crossing of the two nuclei (Cronin effect), which can be described by a Glauber calculation~\cite{lange1}. 

The heavy quarks created in nuclear collisions  propagate  in a strongly interacting and non-static medium, whose properties and evolution is described through hydrodynamical calculations,
 performed with the viscous 2+1 code of
Ref.~\cite{rom1} (for the AA case), or with ECHO-QGP~\cite{ECHO-QGP} (for the pA case, in 2+1 dimensions for simplicity).

While for the initial conditions in AA collisions one can adopt the smooth results provided by an optical-Glauber calculation, in a pA collision the event-by-event fluctuations in the initial state are extremely important, since they are the main source of anisotropic flow in the final state. With a Monte Carlo calculation we have simulated a large number of initial configurations and we have obtained a realistic estimate of the initial average eccentricity. 
In a given initial configuration, with each collision occurring at the transverse coordinate ${\bf x}_i$ (the single nucleons being randomly located with a distribution probability given by a realistic nuclear density), the entropy density in the transverse plane can be defined as
\begin{equation}
 s({\bf x})=\frac{K}{2\pi\sigma^2}
\sum_{i=1}^{N_{coll}}
\,\mbox{exp}\left[-\frac{({\bf x}-{\bf x}_i)^2}{2\sigma^2}\right]
\label{eq:s_dens}
\end{equation}
where $\sigma$ is the ``smearing parameter''  used to  estimate the effective size of the nucleons.
The initial eccentricity, which translate into a non-vanishing elliptic flow~\cite{Niemi}, can be then  evaluated as
 (the brackets denote the average in the transverse plane, with the entropy density in Eq.~\ref{eq:s_dens} as a weight)
\[
\epsilon_2=\frac{\sqrt{\{y^2-x^2\}^2+4\{xy\}^2}}{\{y^2+x^2\}}~.
\]
Since a full event-by-event simulation with our hydro+transport setup would be really demanding, for the huge computing and storage resources required, we evaluate a realistic average background as follows: for a given centrality class we average over all the events of the considered percentile (based on the number of $N_{\scriptscriptstyle \rm{part}}$), after rotating each of them by the event-plane angle $\psi_2$, so that an initial condition representative of the average eccentricity $\epsilon_2$ can be obtained and meaningful predictions for the elliptic flow can be provided.

\begin{figure}
\includegraphics[clip,width=0.4\textwidth]{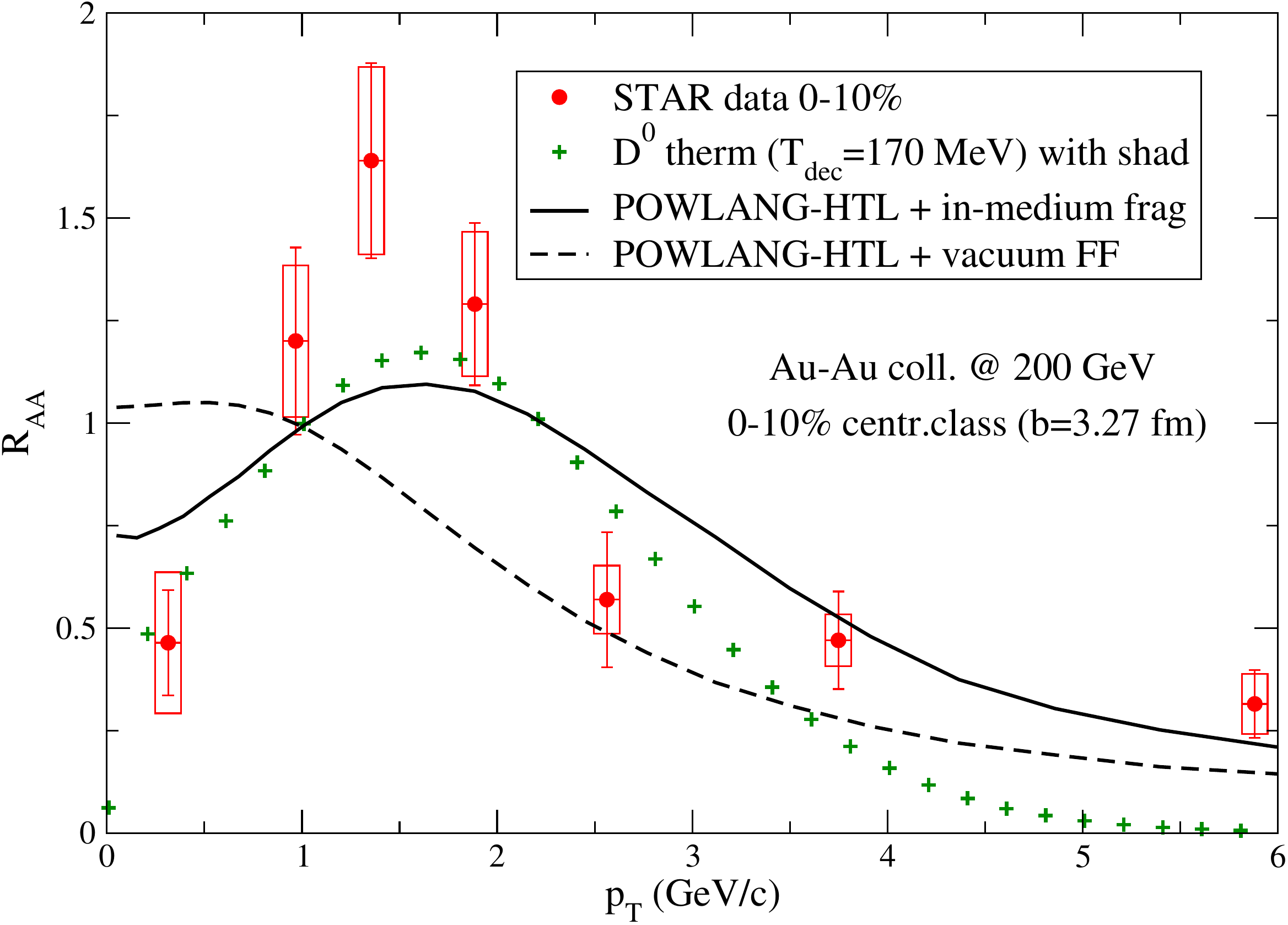} 
\caption{The $R_{AA}$ of $D^0$ mesons in central Au-Au collisions at
  $\sqrt{s_{NN}}\!=\!200$ GeV. POWLANG results obtained with HTL transport
  coefficients and a decoupling temperature $T_{\rm dec}\!=\!170$ MeV are plotted.
Also shown for comparison is the extreme case of full kinetic
thermalization of $D$ mesons. Theory curves are compared to STAR
data \cite{STAR_D0}} 
\label{fig:RAA_D0_1}
\end{figure}

The propagation of the heavy quarks in the plasma  is modeled as a
Brownian motion by employing a relativistic Langevin equation, with 
 the transport coefficients $\kappa_{L/T}(p)$  (representing the average squared
longitudinal/transverse momentum per unit time exchanged by the quark with the 
medium) obtained within a weak-coupling scheme (perturbative QCD and Hard Thermal Loop approximation~\cite{lange0,lange1,lange2})

Finally, after the propagation through the dense me\-dium, the heavy quarks decouple and hadronize.
In the pp case, the hadronization is performed with PYTHIA, while for the nuclear collisions we 
introduce a new simple model to take into account the effect of the thermalized medium.
Once a heavy quark $Q$,
during its stochastic propagation in the fireball, has reached a fluid cell
below the decoupling temperature $T_{\rm dec}$, it is forced to hadronize. One
extracts then a light antiquark $\overline{q}_{\rm light}$ (up, down or strange,
with relative thermal abundances dictated by the ratio $m/T_{\rm dec}$) from a
thermal momentum distribution corresponding to the temperature $T_{\rm dec}$ in
the Local Rest Frame (LRF) of the fluid; the  local fluid
four-velocity $u^\mu_{\rm fluid}$  allows one to boost the momentum of
$\overline{q}_{\rm light}$ from the LRF to the laboratory frame.  
A string is then constructed joining the endpoints given by $Q$ and
$\overline{q}_{\rm light}$ and is then given to PYTHIA 6.4 to
simulate its fragmentation into hadrons (and their final decays). 
In case the
invariant mass of the string is not large enough to allow its decay into at
least a pair of hadrons the event is resampled, extracting a new thermal parton
to associate to the heavy quark.

 \begin{figure}
\includegraphics[clip,width=0.4\textwidth]{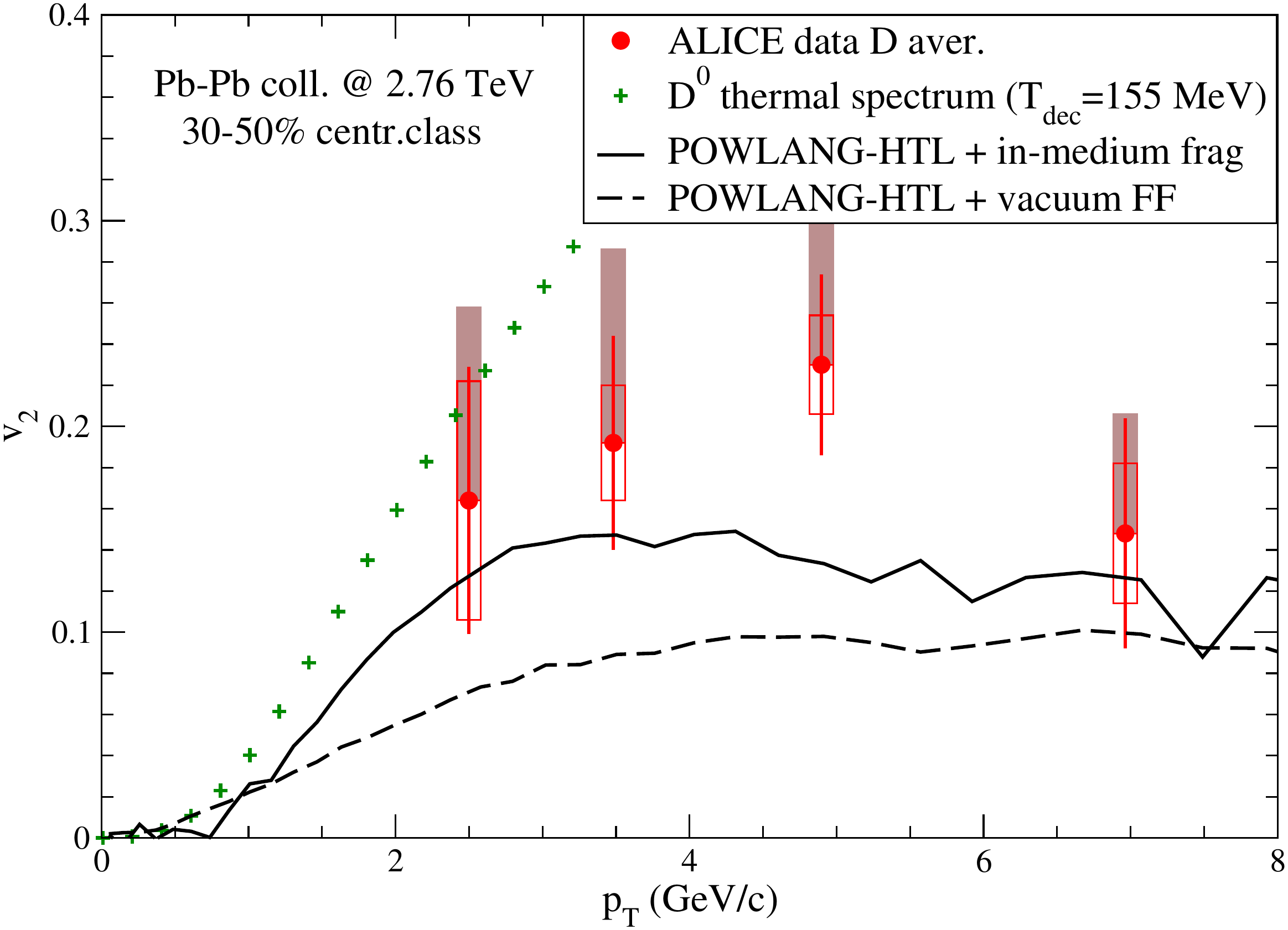} 
\caption{The $v_2$ of $D$ mesons in Pb-Pb collisions at $\sqrt{s_{NN}}\!=\!2.76$
  TeV. POWLANG results (with HTL transport coefficients) with in-vacuum and
  in-medium HQ fragmentation at the decoupling temperature $T_{\rm dec}=155$ MeV
   compared to ALICE data \cite{Abelev:2013lca} in the 30-50\% centrality class and to the limit of
  kinetic thermalization. }\label{fig:v2_D_LHC_transp}
\end{figure}
We would like to stress that this procedure to simulate in-medium fragmentation is new, 
since in our previous works the same mechanism as in pp collisions (i.e. in-vacuum fragmentation) was employed.
With this improvement we can
first of all provide a realistic estimate of the role of the thermal light quarks to explain
peculiar features of the $D$ meson spectra at low and moderate $p_T$; secondly,
the complete information on all the final state particles arising from the
fragmentation of the strings allows us to provide theory predictions for
observables like $D\!-\!h$, $e\!-\!h$, $e^+\!-\!e^-$... correlations to be
compared to existing data and possibly used as a guidance to  future 
measurements.

Notice that further possible interactions in the ha\-dro\-nic phase, which might enhance the elliptic flow,
 are here neglected: we plan to include them in our future work.

\section{Results in pA and AA collisions}
\label{sec:results}

In this section we present a selection of our results, compared with experimental data obtained at RHIC and LHC energies. 

In Fig.~\ref{fig:RAA_D0_1} we show some POWLANG outcomes
for the $R_{AA}$ of $D^0$ mesons in central ($0-10\%$) Au-Au collisions at
$\sqrt{s_{NN}}\!=\!200$ GeV. HTL transport coefficients are employed and the
difference between the two hadronization schemes (here taken to occur at $T_{\rm dec}=170$
MeV\footnote{We notice that the nuclear modification factor $R_{AA}$ is not very sensitive to the value of $T_{\rm dec}$: with a smaller value (155 MeV) the results are quite similar \cite{lange3}}), either with vacuum fragmentation or with in-medium string fragmentation,
are clearly visible: in the second case the radial flow of the light thermal
parton leads to the development of a bump around $p_T\sim 1.5$ GeV in
qualitative agreement with the experimental data. 
Also shown for comparison is the result for the limiting scenario in which 
charmed particles reach full kinetic equilibrium with the medium \cite{lange3}.
Our results are compared to STAR data\cite{STAR_D0}.

In Fig.~\ref{fig:v2_D_LHC_transp}  we address
the $v_2$ of $D$ mesons in semicentral (30-50\%) Pb-Pb collisions at
$\sqrt{s_{NN}}\!=\!2.76$ TeV at the LHC. The effect of the new procedure for
in-medium hadronization through string fragmentation is clearly visible:
while POWLANG outcome with standard in-vacuum fragmentation of charm
largely underpredicts the observed $v_2$, the additional flow acquired from the
light thermal partons move the theory curves with in-medium hadronization closer
to the experimental data measured by the ALICE Collaboration~\cite{Abelev:2013lca}. The value chosen for the decoupling temperature is  $T_{\rm dec}=155$ MeV; we observed that $v_2$ is more sensitive to $T_{\rm dec}$ than other observables, the lower value seeming to be preferred by the data: this agrees with the expectation that the elliptic flow needs more time to fully develop with respect to the quenching of the spectra. 
This plot also shows how a full kinetic thermalization up to large values of $p_T$ is disfavoured by the data.

\begin{figure}
\includegraphics[clip,width=0.4\textwidth]{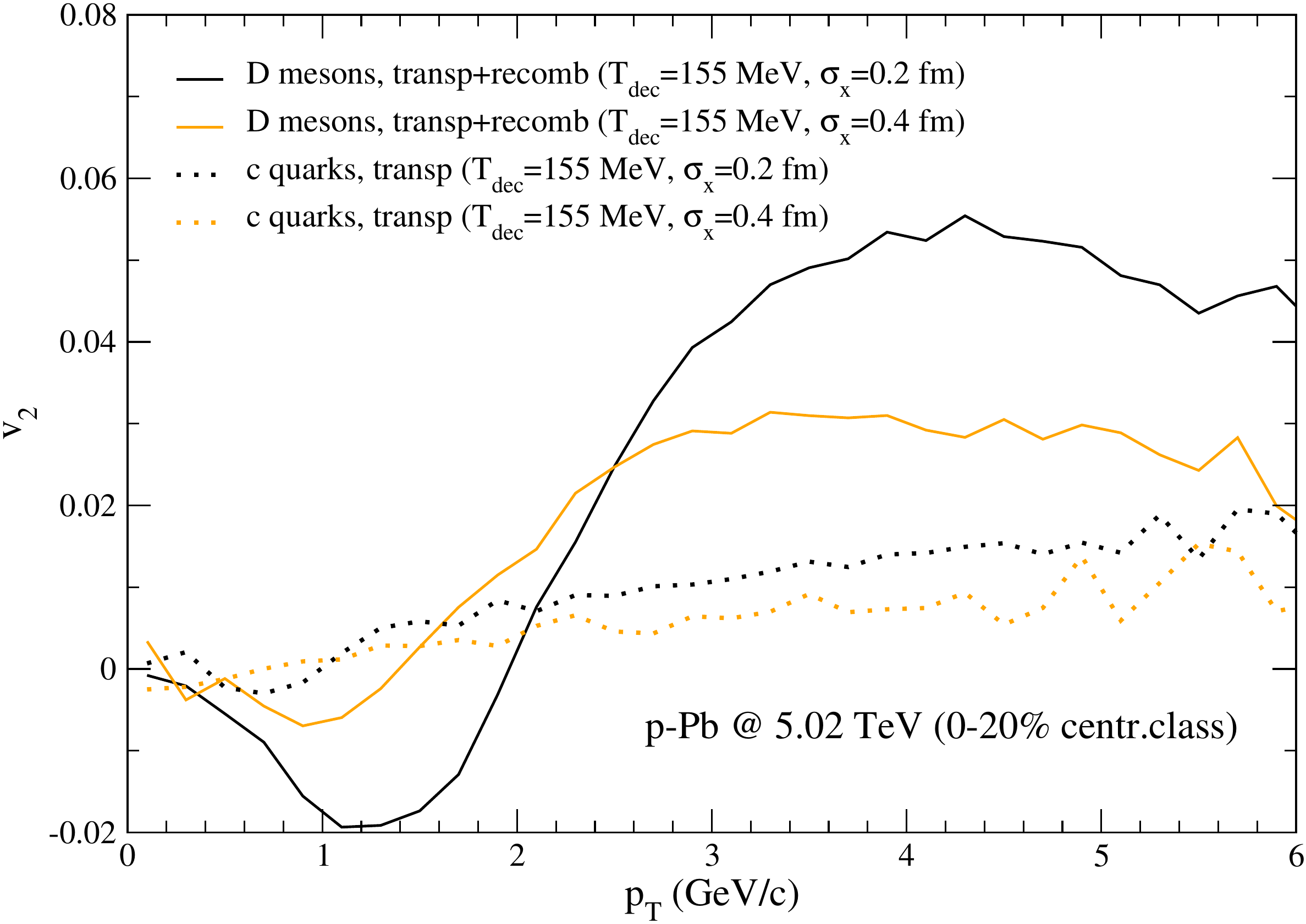}
\caption{The $v_2$ of D mesons in p-Pb collisions at $\sqrt{s_{NN}}\!=\!5.02$
  TeV. POWLANG results (with HTL transport coefficients) with 
  in-medium HQ fragmentation, for $\sigma=0.2$ and $0.4$ fm. The $v_2$ coefficients for c-quarks alone are plotted for comparison.}\label{fig:v2_D_pPb}
\end{figure}

\begin{figure}
\includegraphics[clip,width=0.45\textwidth, height=4.8cm]{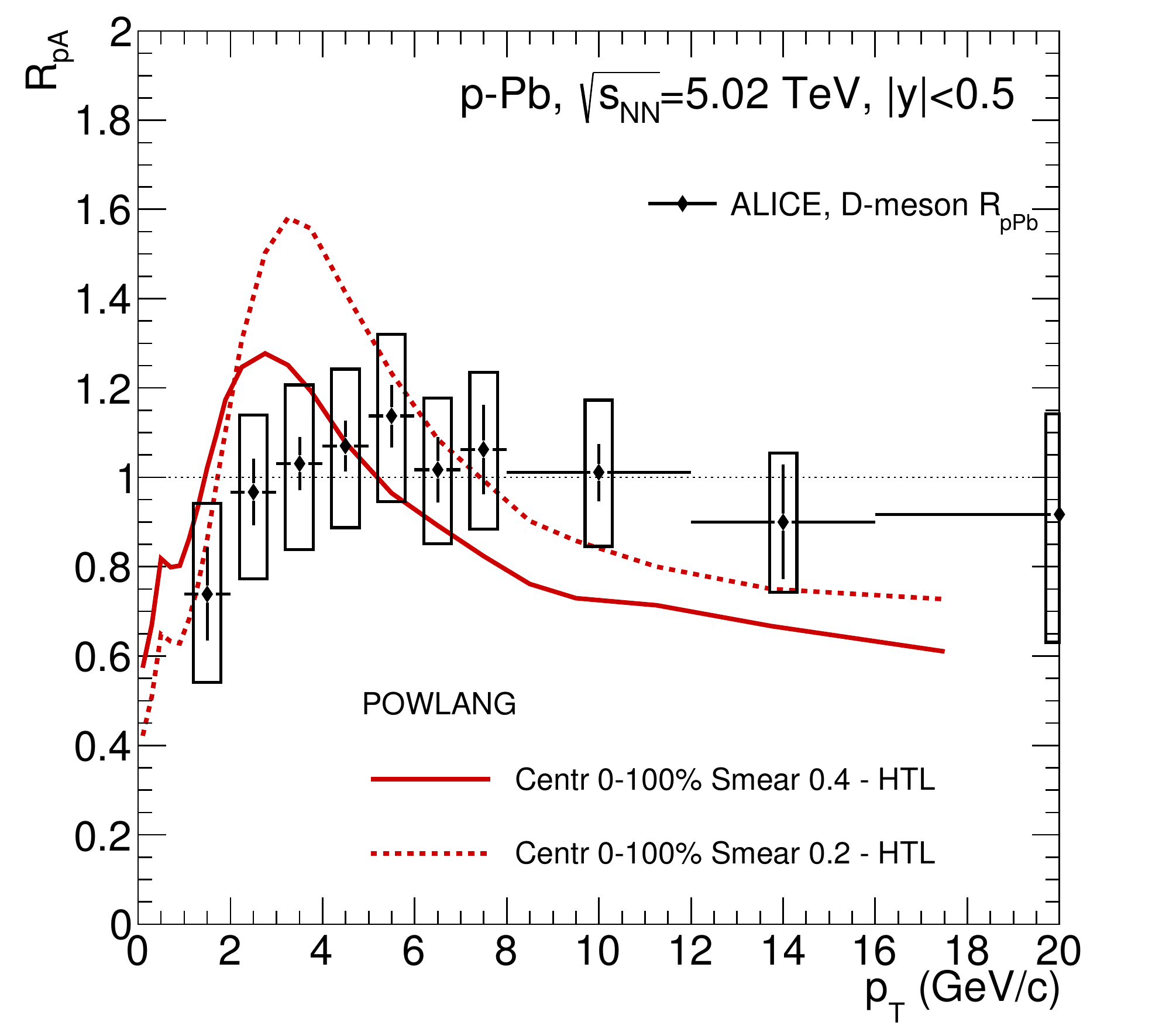}
\caption{The nuclear modification factor of $D$ mesons in p-Pb collisions at $\sqrt{s_{NN}}\!=\!5.02$
  TeV. POWLANG results (with HTL transport coefficients) with 
  in-medium HQ fragmentation, for $\sigma=0.2$ and $0.4$ fm. Experimental data from Ref.~\cite{Abelev:2014hha}}\label{fig:R_D_pPb}
\end{figure}

Finally, we show some results for pA collisions. In Fig.~\ref{fig:v2_D_pPb} we show the elliptic flow parameter $v_2$ for D mesons in p-Pb collisions at $\sqrt{s_{NN}}=5.02$ TeV. We have plotted the results for two values of the smearing parameter $\sigma$. The dotted lines show the $v_2$ for $c$-quarks alone: it is evident the importance of the contribution of the light quarks to the collective flow of the charmed mesons. In  Fig.~\ref{fig:R_D_pPb} we plot the nuclear modification factor for $D$ mesons: we note that the POWLANG setup predicts a peak at $p_T\sim 3\div 4$ GeV, with a systematic uncertainty arising from the value of
the initial smearing parameter. 

We plan to extend this study by considering trasport coefficients evaluated on the lattice and to apply
the same setup  also to d-Au collisions, experimentally measured at RHIC.

\section{CONCLUSIONS}

The simple model to describe heavy quark
hadronization in the presence of a hot deconfined medium (a Quark-Gluon Plasma)
has considerably improved the agreement of the POWLANG results with the experimental data at RHIC and LHC energies.  
In particular, results for the  nuclear modification factors and the elliptic flow of D mesons, for pA and AA collisions, have been presented.

With this setup, it is also possible to study  heavy-flavour correlations in heavy-ion collisions,
 allowing to get additional information on the heavy quark interaction
with the medium.


\begin{thebibliography}{99}

 
 \bibitem[Beraudo et al. (2009)]{lange0} 
  A. Beraudo, A. De Pace, W.M. Alberico, A. Molinari,
                    Nucl. Phys. A {\bf 831}, 59 (2009)
 
 \bibitem[Alberico et al. (2009)]{lange1}  
  W.M. Alberico, A. Beraudo, A. De Pace, A. Molinari, M. Monteno, M. Nardi, F. Prino, 
                    Eur. Phys. J. C {\bf 71}, 1666 (2011)
 \bibitem[Alberico et al. (2013)]{lange2}   
W.M. Alberico, A. Beraudo, A. De Pace, A. Molinari, M. Monteno, M. Nardi, F. Prino, M. Sitta, 
 		  Eur.Phys.J. {\bf C73},  2481 (2013)
 
 \bibitem[Beraudo et al.(2014)]{lange3} 
   A.~Beraudo, A.~De Pace, M.~Monteno, M.~Nardi and F.~Prino, 
Eur.Phys.J. {\bf C75}, 121 (2015) 3
 \bibitem[Frixione et al.(2007)]{POW} S. Frixione, P. Nason, G. Ridolfi,
                  JHEP {0709} (2007) 126.
 \bibitem[Alioli et al.(2010)]{POWBOX} S. Alioli, P. Nason, C. Oleari and E. Re,
 JHEP {1006} (2010) 043.
 
 \bibitem[Sjostrand et al.(2006)]{PYTHIA} T. Sjostrand, S. Mrenna and P.Z. Skands,
 JHEP 0605 (2006) 026.
 
\bibitem[Bjelogrli\'c (2014)]{sandro} S. Bjelogrli\'c (ALICE Collab.), Nucl. Phys  A {\bf 931}, 563 (2014)
\bibitem[Rossi (2014)]{rossi} A. Rossi  (ALICE Collab.),
Nucl. Phys A {\bf 932}, 51 (2014)

\bibitem[Eskola et al.(2009)]{eps}      K.J. Eskola, H. Paukkunen, C.A. Salgado,
                    J. High Energy Phys. {\bf 0904}, 065 (2009)

\bibitem[Romatschke and Romatschke(2007)]{rom1}  P. Romatschke, U. Romatschke,
                    Phys. Rev. Lett. {\bf 99}, 172301 (2007).
\bibitem{ECHO-QGP} ECHO-QGP project:~{\tt http://theory.fi.infn.it/echoqgp}
\bibitem{Niemi}
  H.~Niemi, K.~J.~Eskola and R.~Paatelainen,
  arXiv:1505.02677 [hep-ph].
\bibitem[STAR(2014)]{STAR_D0} L.Adamczyk \emph{et al.}
 (STAR Collab.), { arXiv:1404.6185 [nucl-ex]}.

\bibitem[ALICE(2012)]{ALICE_Abelev}
 ALICE Collaboration,
   JHEP {\bf 1211} (2012) 065.

\bibitem[ALICE(2013)]{delValle}
   Z.~Conesa del Valle (ALICE Collaboration),
   Nucl.\ Phys.\ A {\bf 904-905} (2013) 178c.
\bibitem[ALICE(2013)]{Abelev:2013lca}
   B. Abelev  (ALICE Collaboration),
   Phys.\ Rev.\ Lett.\  {\bf 111} (2013) 102301.
 
\bibitem{Abelev:2014hha}
  B.~B.~Abelev  (ALICE Collaboration),
  Phys.\ Rev.\ Lett.\  {\bf 113} (2014) 23,  232301.


%
%
%
%
%
%
%
%
%
 \end{thebibliography}
\end{document}